\documentclass[aps,pra,reprint,groupedaddress,amsfonts,amssymb,amsmath]{revtex4-1}
\pdfoutput=1
\usepackage{amsmath,amssymb}
\usepackage{lipsum}
\usepackage{amsfonts}
\usepackage{ graphicx } % Include figure files
\usepackage{ bm } % bold math
\usepackage{ color }
\usepackage{float}
\usepackage[T1]{fontenc}

\newcommand{\be}{\begin{equation}}
\newcommand{\ee}{\end{equation}}
\newcommand{\bea}{\begin{eqnarray}}
\newcommand{\eea}{\end{eqnarray}}

\newcommand{\mele}[3]{\left\langle #1 \middle | #2 \middle | #3 \right\rangle} %%% <1|a|3> matrix element in bra-ket notation
\newcommand{\ket}[1]{\left | #1 \right\rangle}

\newcommand{\avg}[1]{\left\langle #1 \right\rangle} %% expectation value <1>

\newcommand{\Kpi}{K_{\pi/2}}
\newcommand{\Keps}{K_{-\epsilon\pi}}

\begin{document}
\title{Discrete time crystal in globally driven interacting quantum systems without disorder}

\author{Wing Chi Yu$^{1}$}
\email{wcyu.physics@gmail.com}
\author{Jirawat Tangpanitanon$^{1}$}
\email{a0122902@u.nus.edu}
\author{Alexander W. Glaetzle$^{1,2}$}
\author{Dieter Jaksch$^{2,1}$}
\author{Dimitris G. Angelakis$^{1,3}$}
\email{dimitris.angelakis@gmail.com}

\affiliation{$^{1}$Centre for Quantum Technologies, National University of Singapore, 3 Science Drive 2, Singapore 117543, Singapore}
\affiliation{$^{2}$Clarendon Laboratory, University of Oxford, Parks Road, Oxford OX1 3PU, United Kingdom}
\affiliation{$^{3}$School of Electronic and Computer Engineering, Technical University of Crete, Chania, Crete, 73100 Greece}

\date{ \today }

\begin{abstract}
Time crystals in periodically driven systems have initially been studied assuming either the ability to quench the Hamiltonian between different many-body regimes, the presence of disorder or long-range interactions. Here we propose the simplest scheme to observe discrete time crystal dynamics in a one-dimensional driven quantum system of the Ising type with short-range interactions and no disorder. The system is subject only to a periodic kick by a global magnetic field, and no extra Hamiltonian quenching is performed. We analyze the emerging time crystal stabilizing mechanisms via extensive numerics as well as using an analytic approach based on an off-resonant transition model. Due to the simplicity of the driven Ising model, our proposal can be implemented with current experimental platforms including trapped ions, Rydberg atoms, and superconducting circuits.
\end{abstract}


\maketitle

%%\textit{Introduction.---} 
\begin{figure}[t]
	\centering
	\includegraphics[width=8cm]{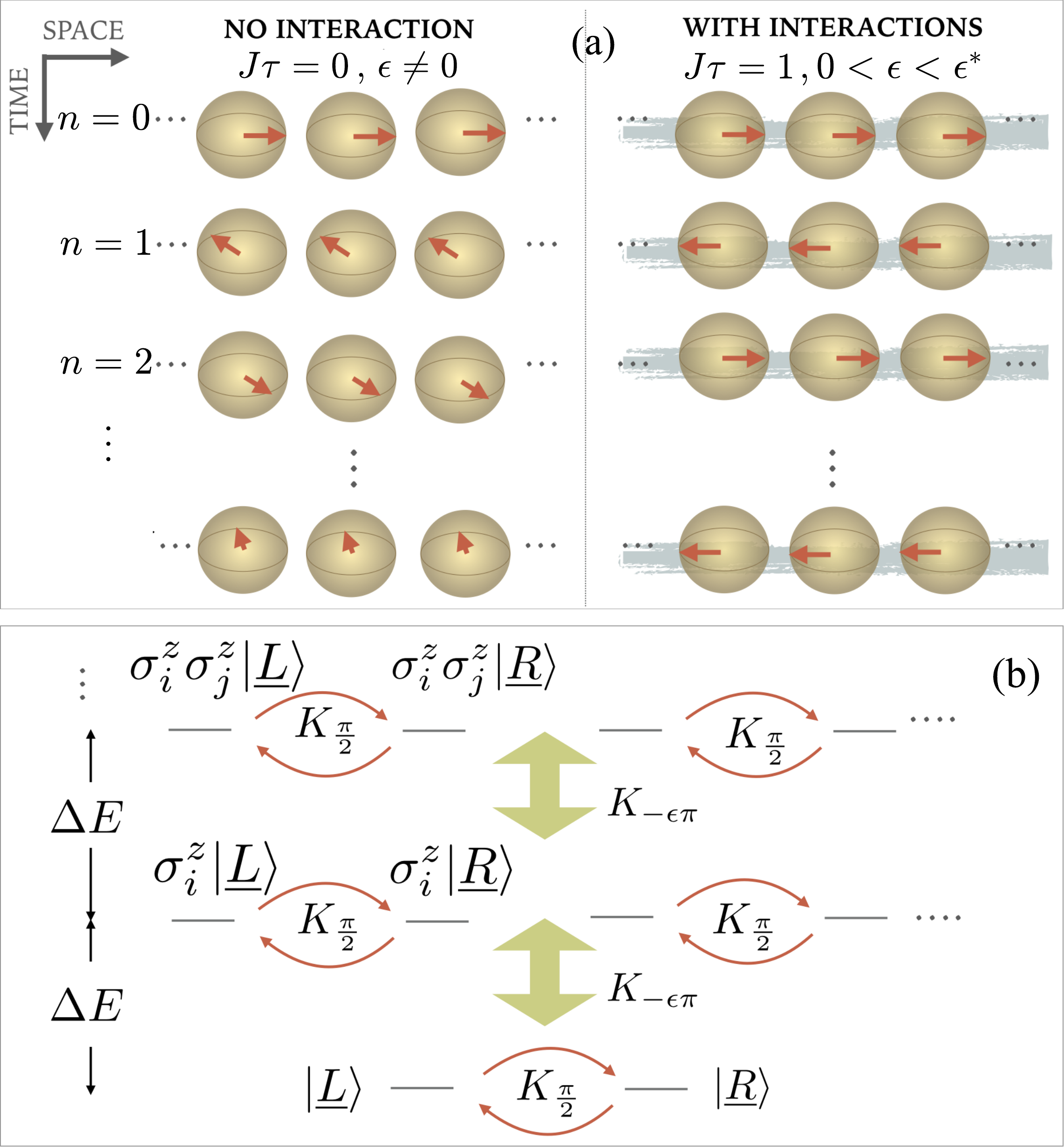}
	\caption{ (a) Schematic diagram showing the dynamics of the individual spins in our model for a small perturbation $\epsilon$ in the spin flip. The presence of interaction helps to synchronize the spins. (b) The energy spectrum of the NN Ising model for $h=0$ and the driving scheme. The energy levels are evenly spaced with a gap $\Delta E=2J$. The kick operator $K_{\frac{\pi}{2}}$ only connects two eigenstates within the same degenerate manifold, while the errors in the kick operator $K_{-\epsilon \pi}$ may couple different manifolds. However for the case of off-resonant driving as in our protocol where the driving frequency $\omega_d=2\pi/\tau$ is far from the energy gap $\Delta E=2J$, the effect of $K_{-\epsilon \pi}$  is switched off for $\epsilon \ll \epsilon^*$ and the DTC is robust. }\label{fig1}
\end{figure}

\section{Introduction}
In 2012, Wilczek proposed the idea of quantum time crystals which spontaneously break the continuous time translation symmetry \cite{Wilczek2012}. He suggested that a ring of interacting bosons prepared in the ground state can switch to a periodic motion in time if the magnetic flux through the ring is properly chosen. However, a no-go theorem later pointed out that such a time crystal phase is forbidden in equilibrium \cite{Bruno2013,Watanabe2015}. Alternatively, Sacha first proposed to search for time crystal dynamics in periodically driven systems \cite{Sacha2015} which was further concretized by Khemani \textit{et al.} \cite{Khemani2016} and Else \textit{et al.} \cite{Else2016} respectively studying many-body models. In the presence of strong disorder, the system is many-body localized (MBL) and does not absorb heat from the drive. In this MBL regime, the system can oscillate with a period which is different from the drive's without thermalizing to an infinite temperature. Such a phase is known as a discrete (or Floquet) time crystal (DTC) to emphasize the discreteness and to differentiate from the original proposal by Wilczek. Subsequent theoretical and numerical studies have demonstrated the existence of DTC in various disordered Floquet systems \cite{Keyserlingk2016,Yao2017,Mierzejewski2017,Lazarides2017}.

Recently, DTCs have been observed in various experiments with trapped ions \cite{Zhang2017}, spatial crystals ammonium dihydrogen phosphate $\rm{NH_4H_2PO_4}$ \cite{Rovny2018,Rovny2018b}, and nitrogen-vacancy centers in diamond \cite{Choi2017} in the presence of disorder or long-range interactions. While in Ref. \cite{Zhang2017}, DTCs were realized in the MBL phase, the disorder in Refs. \cite{Choi2017,Rovny2018,Rovny2018b} was insufficient for reaching the MBL regime. This triggered a search for DTCs that are not protected by MBL \cite{WWHo2017,Abanin2017,Kucsko2017,TSZeng2017,PRX2017,Torre2017,Pal2018,PRX2017,BHuang2018,Russomanno2017,Giergiel2018}. Driven many-body systems without disorder that exhibit a DTC have been proposed for quenched Hamiltonian with short-range interactions in cold atoms \cite{BHuang2018},  in two dimensions or higher \cite{PRX2017} , in the regime with all-to-all spin interactions \cite{Russomanno2017} and ultracold atoms bouncing on an
oscillating atom mirror \cite{Giergiel2018}.

In this work, we study a DTC in a simple periodically driven one-dimensional Ising quantum chains with finite-range two-body interactions and no disorder. In contrast to Ref. \cite{BHuang2018} where the driving protocol involves quenching between two many-body Hamiltonians, which is experimentally challenging, our drive only consists of delta kicks generated by a global magnetic field that periodically applies a $\pi/2-$pulse to each spin.

We find that spin-spin interactions, regardless of their range, help to stabilize the time crystal with a period doubling against small errors in the driving protocol. We analyze the stabilizing mechanism by providing a perturbative model that analyzes the effect of unwanted off-resonant transitions to other undesired states created by errors in the driving protocol. Our setup can be experimentally implemented in all quantum technologies platforms that can realize the quantum Ising model, including trapped ions \cite{Britton2012,Islam2013,Jurcevic2014,Richerme2014,Bohnet2016}, Rydberg atoms \cite{Labuhn2016,Zeiher2017}, superconducting circuits \cite{dwave}, and solid state $\rm{LiHoF_4}$ \cite{Silevitch2010}.

%\section{The setup}

\begin{figure*}[t]
	\centering
	\includegraphics[width=14cm]{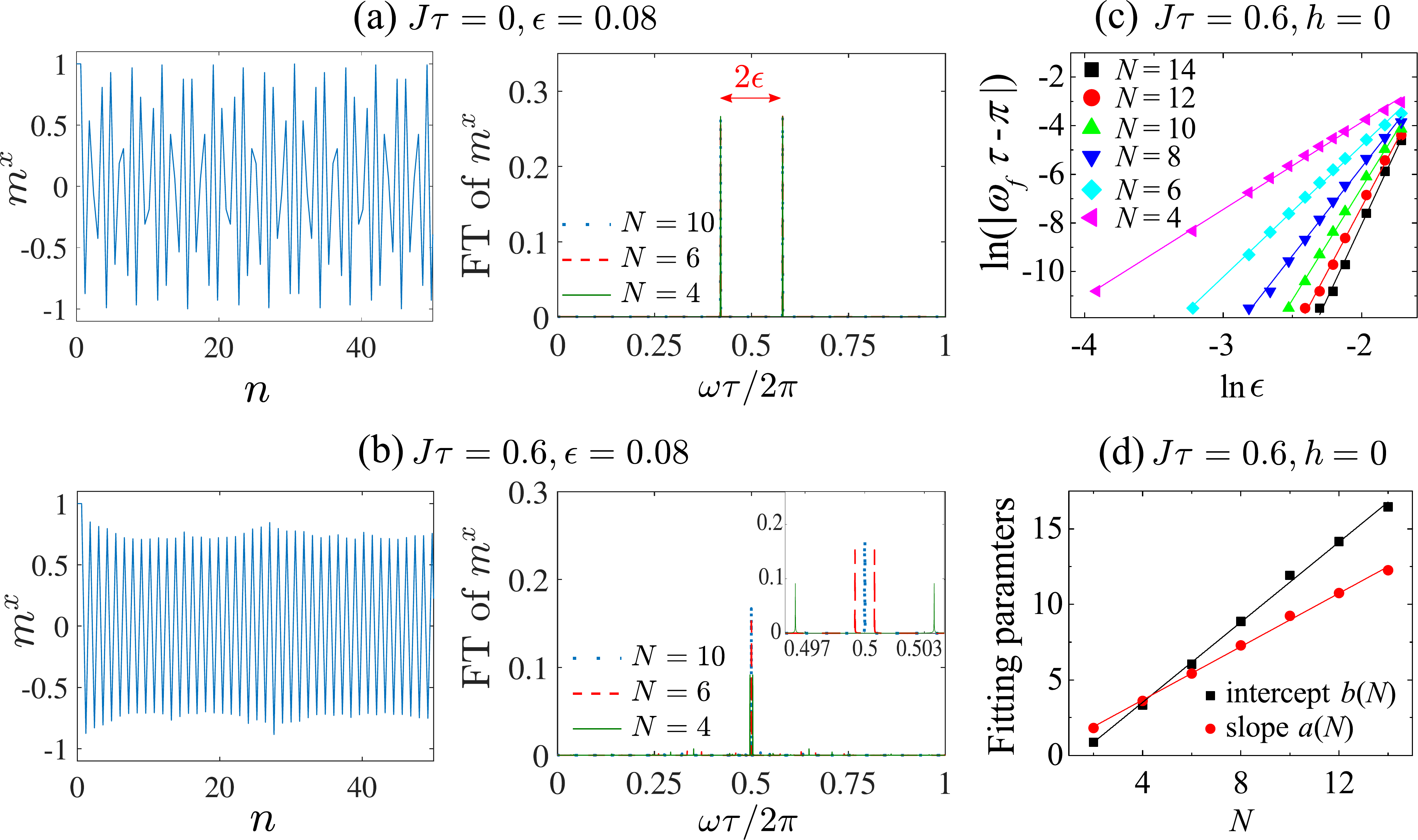}
	\caption{ [(a) and (b)] Stroboscopic magnetization $m^x(n)$ for $N=10$ and Fourier transforms of $m^x(n)$ for various system sizes $N$. Here $\epsilon=0.08, h=0$. In the absence of spin-spin interaction as shown in (a), the Fourier peaks for various $N$ coincide. The splitting of the two peaks equals $2\epsilon$ for all $N$ while it scales approximately as $(\epsilon/\epsilon^*)^{m_aN}$ from the scaling analysis in (c) and (d) in the presence of interaction. (c) Fourier peak splitting as a function of $\epsilon$ for $h=0,J\tau=0.6$. The straight lines show the linear fits for various system sizes. (d) The linear fitting parameters of (c) as a function of $N$. Straight lines show the linear fitting of the data points.}\label{fig2}
\end{figure*}

\begin{figure*}[t]
	\centering
	\includegraphics[width=13.2cm]{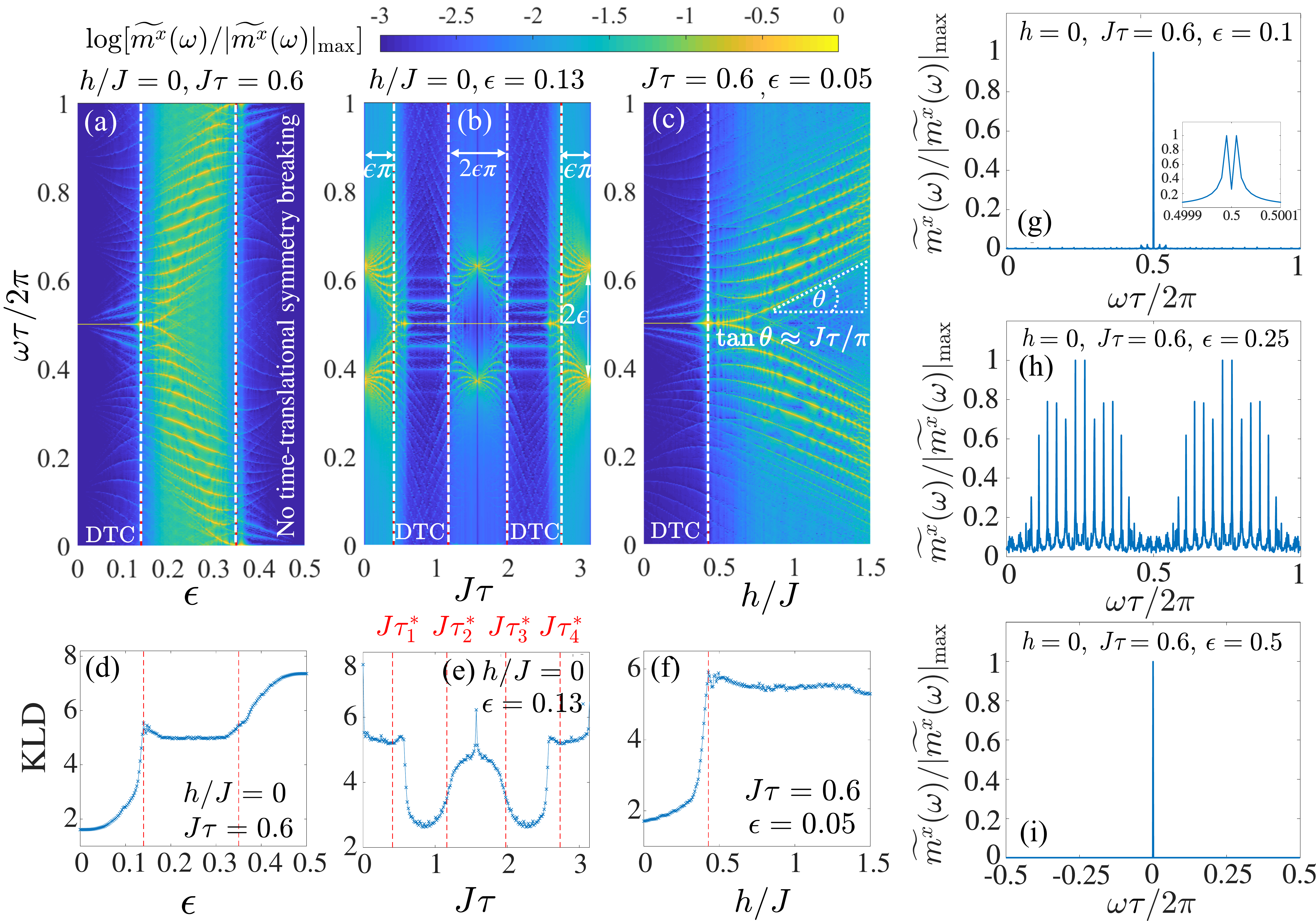}
	\caption{ [(a)-(c)] Color map of the Fourier spectrum of the stroboscopic magnetization in $x$ direction with $\epsilon$, $J\tau$ and $h/J$ as the driving parameter respectively. Here $N=14$ and the Fourier transform is performed over 1000 periods of the drive. The amplitude of the Fourier peaks is divided by the maximum amplitude for better visualization. [(d)-(f)] KL divergence of the Fourier spectrum in the corresponding upper panel.  The plateaus around KLD$\sim 2$ indicate the DTC phase and the vertical dash lines show the approximated phase boundary. [(g)-(i)] Cuts of the Fourier spectrum color map in (a) for $\epsilon$ from three different phases. (g) $\epsilon=0.1$ and the system is in DTC phase. Inset shows a zoom-in of the main peaks around $\omega\tau=\pi$ (the Fourier transform is carried out over $10^5$ periods here). The two main peaks are separated by $(\epsilon/\epsilon^*)^{m_aN}$ in this phase. (h) $\epsilon=0.25$ and there is no prominent peak observed. (i) $\epsilon=0.5$ and the system oscillate with the drive. One prominent peak is observed at $\omega=0$. The spectrum is folded into $[-\pi,\pi)$ for better visualization of the main peak. }\label{fig3}
\end{figure*}

\begin{figure}[b]\centering
	\includegraphics[width=8.5cm]{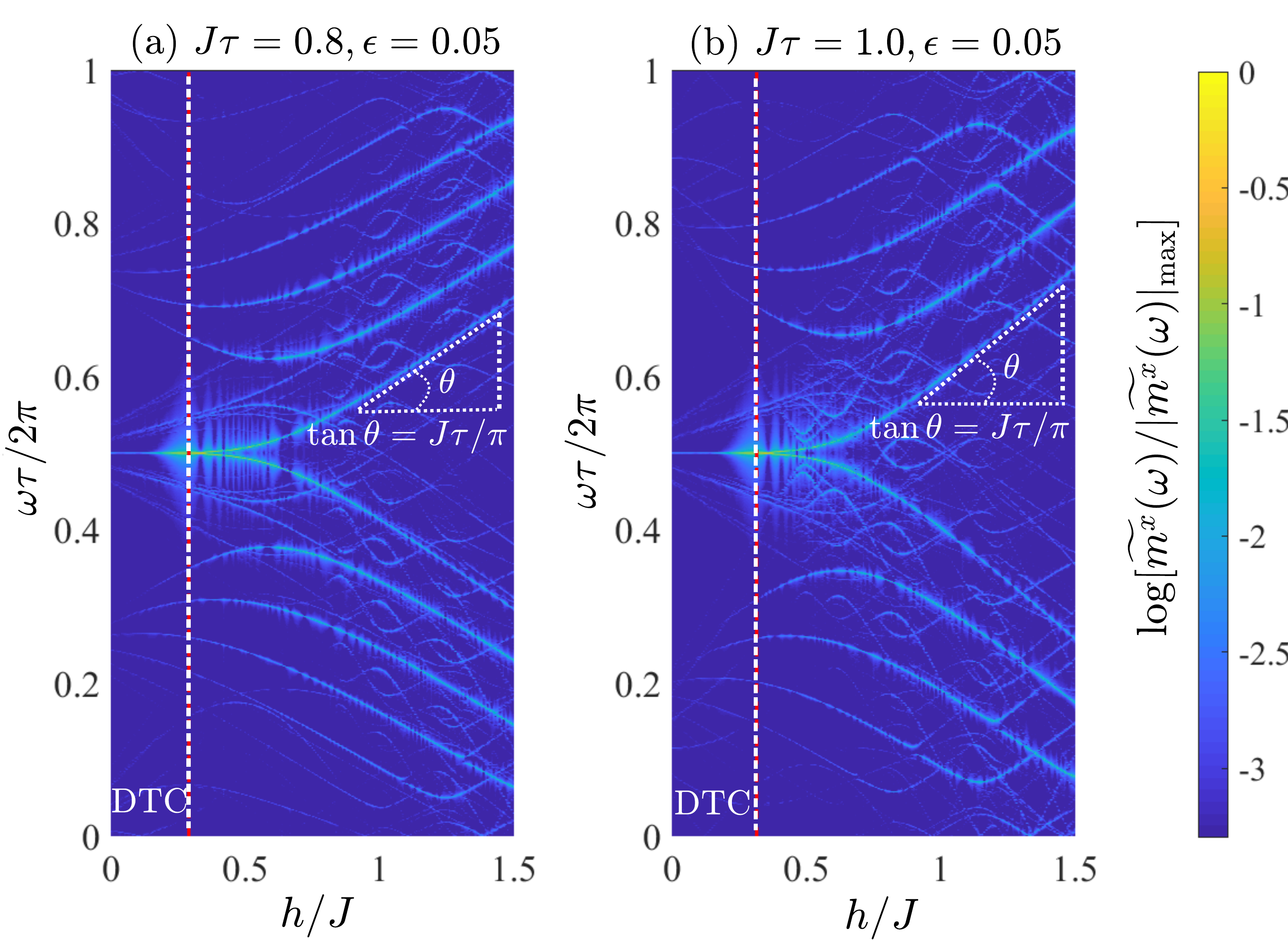}
	\caption{\label{fig4} [(a) and (b)] Color map of the Fourier spectrum of the stroboscopic magnetization in the $x$ direction with $h/J$ as the driving parameter for various $J\tau$. Here $N=8$ and the Fourier transform is performed over 1000 periods.}
\end{figure}

\section{The system, state preparation and driving protocol} 
We consider the dynamics of the Ising model in a transverse-field described by the Hamiltonian
\begin{equation}
H_0 = -J\sum_{i}\sigma_i^x\sigma_{i+1}^x-h\sum_{i}\sigma_i^z
\label{eq:H_LRI}
\end{equation}
under a periodic delta kick in the absence of disorder. Here $\sigma_i^{\kappa}$ ($\kappa=x,y,z$) is the Pauli matrix operator at site $i$, $h$ is the strength of the transverse field and $J$ is the spin-spin coupling strength. %Unless otherwise specified, the units of energy and time are set to $J$ and $1/J$ respectively.  
%With $\alpha$ we characterize the range of interaction. For examples, $\alpha=1$ refers to the Coulomb interactions, $\alpha=3$ is the dipole-dipole interaction, and $\alpha=6$ corresponds to the van der Waals interaction. In the limit of $\alpha\rightarrow\infty$, the Hamiltonian reduces to the usual nearest-neighbor quantum Ising model which 
The model exhibits a quantum phase transition at $h_c=J$ in the thermodynamic limit. For $h>h_c$, the system's ground state is a quantum paramagnet while for $h<h_c$, the system breaks the $\mathbb{Z}_2$ symmetry spontaneously and becomes a ferromagnet \cite{Sachdev2000}. %,Jaschke2017,ZZhu2018}. %For a general value of $\alpha>1$, %which will be our focus in this work,
%previous studies via the truncated Jordan-Wigner transformation have shown that the critical point occurs at $h_c=\zeta(\alpha)$, where $\zeta$ is the Riemann Zeta function \cite{Jaschke2017}.

The initial state $\ket{\Psi(0)}$ is prepared in one of the two ferromagnetic ground states of $H_0$ with $h=0$. These ground states are simply the product states $|\underline{R}\rangle\equiv \otimes_i\ket{\rightarrow}_i$ and $|\underline{L}\rangle\equiv\otimes_i\ket{\leftarrow}_i$, where $\ket{\rightarrow}_i$ and $\ket{\leftarrow}_i$ are eigenstates of $\sigma^x_i$. After evolving the system with $H_0$ for a time $\tau$, we apply a delta kick
\begin{equation}
K_{\phi}=\exp\left[{-i\phi\sum_i\sigma_i^z}\right],
\end{equation}
which rotates the spins about the $z$ axis by an angle $\phi=\pi(1/2-\epsilon)$, where $\epsilon$ is a perturbation. The procedure is then repeated. The time evolution operator over one period is thus
\begin{equation}
U = K_{\phi}U_0 = K_{-\epsilon\pi}K_{\pi/2}U_0,
\label{eq:UF}
\end{equation}
where $U_0=\exp\left[{-iH_0\tau}\right]$ and $\hbar=1$. For $\epsilon=0$ and $h=0$, the kick operator will just flip the spins at every time $n\tau$ and the system returns to the initial state at every $2\tau$. As will be shown below, the spin-spin interaction in $H_0$ can `correct' the imperfect spin flip for a small but finite $\epsilon$ (Fig. \ref{fig1}(a)), causing the formation of the time-crystal \cite{LMG}.

To observe the DTC, we measure the total magnetization in the $x$ direction at every period, i.e.,
\bea
m^x(n) = \frac{1}{N}\langle\Psi(n) | \sum_i\sigma_i^x |\Psi(n)\rangle,
%m^x(n) = \frac{1}{N}\mele{\Psi(n)}{ \sum_i\sigma_i^x }{\Psi(n)},
\label{eq:order_para}
\eea
where $N$ is the number of spins in the system and $\ket{\Psi(n)}=U^n\ket{\Psi(0)}$ is the wavefunction of the system just before the $n$-th kick. We will show that, under an off-resonant driving condition, $m^x(n)$ fulfills the following criteria for the DTC in the thermodynamic limit \cite{BHuang2018,Russomanno2017}. (1) Time-translational symmetry breaking: $m^x(n+1)\ne m^x(n)$. (2) Rigidity: $m^x(n)$ shows a fixed oscillation period without fine-tuned Hamiltonian parameters. (3) Persistence: the oscillations must persist for infinitely-long times. Thus the Fourier transform of $m^x(n)$ has a pronounced peak at $\pi$.

%%%%%%%%%%%%%%%%%%%%%%%%%%%%%%%%%%%%%%%%%%%

\section{Effective analytic model}To understand the DTC dynamics in our model, let us first consider the trivial case with $J\tau=h=0$ where all spins are disconnected and start with an initial state $|\underline{R}\rangle$. The state after $n$ driving periods is simply $|\Psi(n)\rangle = K_{\phi}^n|\underline{R}\rangle=K_{ n\phi}|\underline{R}\rangle$ with the magnetization $m^x(n)=(-1)^n\left[2\cos^2(\epsilon\pi n)-1\right]$. Hence, the Fourier spectrum of $m^x(n)$ has two peaks at $\pi\pm 2\epsilon \pi$, as depicted in Fig. \ref{fig2}(a). This is not a time crystal since the positions of the peaks depend on $\epsilon$ regardless of the system size.

When the interaction is switched on ($J\tau\ne 0,h=0$), the above situation changes dramatically. As will be shown, when the drive is off-resonant, the two main peaks will be separated by a distance proportional to $(\epsilon/\epsilon^*)^{m_aN}$ for a critical value $\epsilon^*$ and a positive constant $m_a\sim O(1)$, as depicted in Fig. \ref{fig2}(b). The main peaks' separation will converge to zero as $N\to\infty$ for $\epsilon<\epsilon^*$.
To see this peak merging, let us consider the state after $n$ driving periods $|\Psi(n)\rangle = (K_{\frac{\pi}{2}}K_{-\epsilon \pi}U_0)^n|\underline{R}\rangle$. The driving scheme is depicted in Fig. \ref{fig1}(b). Since $\langle \underline{L} |K_{\frac{\pi}{2}}|\underline{R}\rangle=1$, the kick operator $K_{\frac{\pi}{2}}$ only flips the two ground states $|\underline{R}\rangle$ and $|\underline{L}\rangle$. It does not connect them to the excited states.
%The time crystal phase can exist without neither long-range interactions nor disorders as long as (1) the system has two gapped $\mathbb{Z}_2$ symmetry broken ground states.
%When $\epsilon=0$, these two ground states form a degenerate subspace as depicted in Fig. \ref{fig1}(b).
The operator $K_{-\epsilon \pi}$, on the other hand, generates transitions to the excited states. In first order in $\epsilon$, it does not connect the two ground states. We can see this by approximating $K_{-\epsilon \pi}$ to first order in $\epsilon$ by $K_{-\epsilon \pi}\approx 1+i\epsilon\pi\sum_{i=1}^{N}\sigma^z_i$. Since $K_{-\epsilon \pi}$ only flips one spin, it follows that $\langle \underline{L}|K_{-\epsilon \pi}|\underline{R}\rangle\rightarrow 0$ as $\epsilon\rightarrow 0$. However, the kick operator couples the ground state to the first excited states $\ket{j}=\sigma^z_j\ket{\underline{R}}$ for $j\in\{1,2,\cdots N\}$ as $\mele{j}{K_{-\epsilon\pi}}{\underline{R}}\sim\epsilon$.  
%Consider for example $\tau=0.6$ which corresponds to the frequency $\omega_d=2\pi/\tau=10.47$ in our simulation,
If the driving frequency $\omega_d=2\pi/\tau$ is much larger than the energy gap $\Delta E$ of $H_0$, the corresponding transition to the excited states is too far off-resonant to get significantly populated. Hence $K_{-\epsilon \pi}$ is effectively switched off, as will be confirmed by exact diagonalization below.

With the above conditions, the Fourier spectrum of $m^x(n)$, defined as $\widetilde{m^x}(\omega)$, will show a main peak at $\omega\tau=\pi$ and side peaks of height $\sim\epsilon^2$ \cite{SI_perturbation}. When $\epsilon$ becomes large, one has to take into account higher order terms in the expansion of $K_{-\epsilon\pi}$. The $N$-th order term, in particular, couples the two ground states $|\underline{L}\rangle$ and $|\underline{R}\rangle$, leading to the splitting of the main peak $\delta\omega=|\omega_f\tau-\pi|\approx(\epsilon/\epsilon^*)^{m_aN}$ with $m_a=1$, and $\omega_f$ is the main peak's frequency.

%%%%%%%%%%%%%%%%%%%%%%%%%%%%%%%%%%%%%%%
\section{Discrete time crystal with nearest-neighbour interactions}

\begin{figure}[t]\centering
	\includegraphics[width=8.5cm]{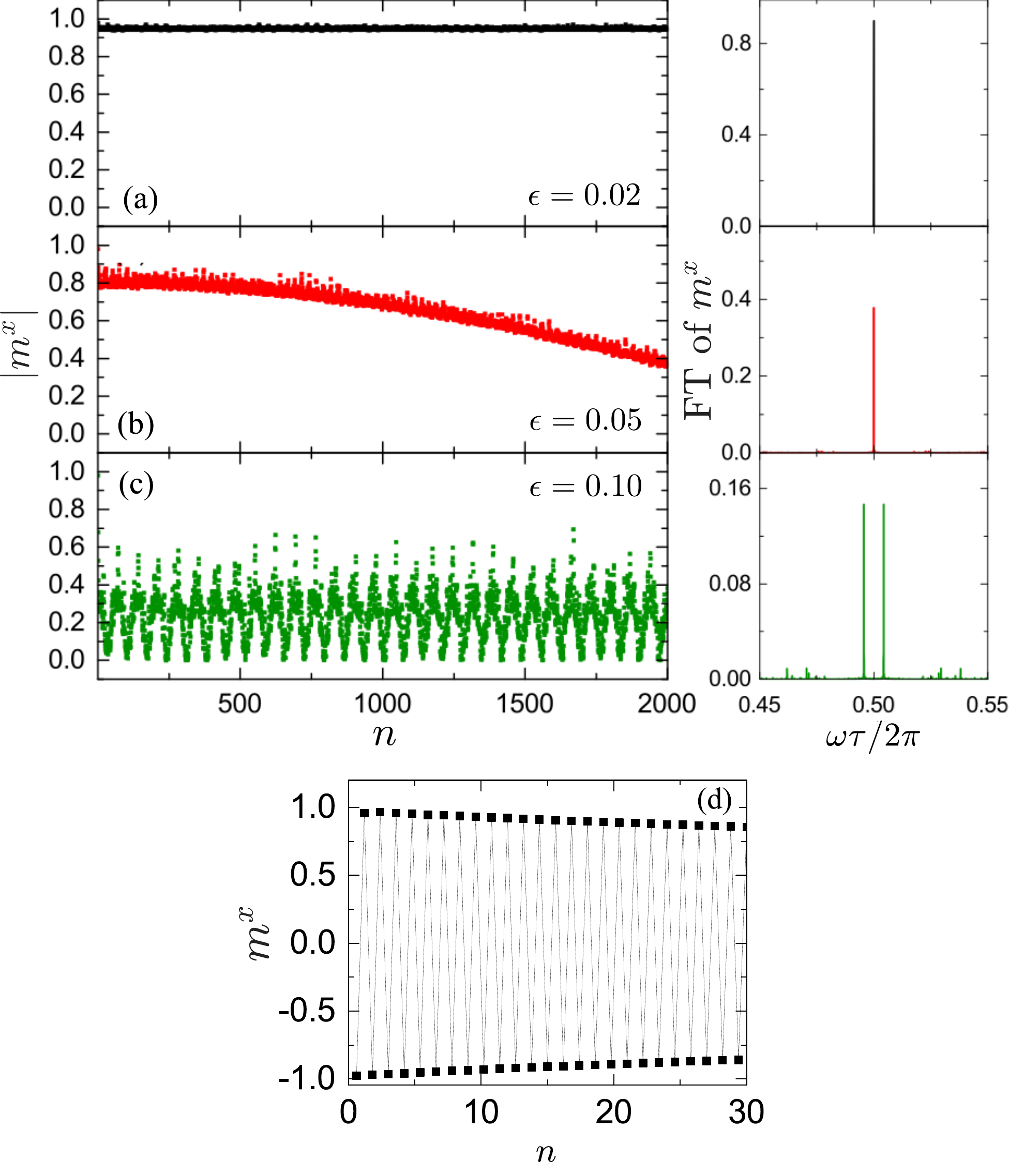}
	\caption{\label{fig5} [(a)-(c)] Stroboscopic magnetization as a function of time for various values of $\epsilon$ and the corresponding Fourier spectrum performed over $10^5$ kicking periods for $N=14,h/J=0.32,J\tau=0.6$. The initial state is prepared in one of the ferromagnetic ground states of $H_0$ with $h/J=0.32$.  (d) Stroboscopic magnetization from TEBD simulation \cite{Dieter2017} for $N=80,\epsilon=0.02, J\tau=0.6, h/J=0.32$. The lines are just guide to the eyes.}
\end{figure}

\begin{figure}[t]
	\centering
	\includegraphics[width=8.5cm]{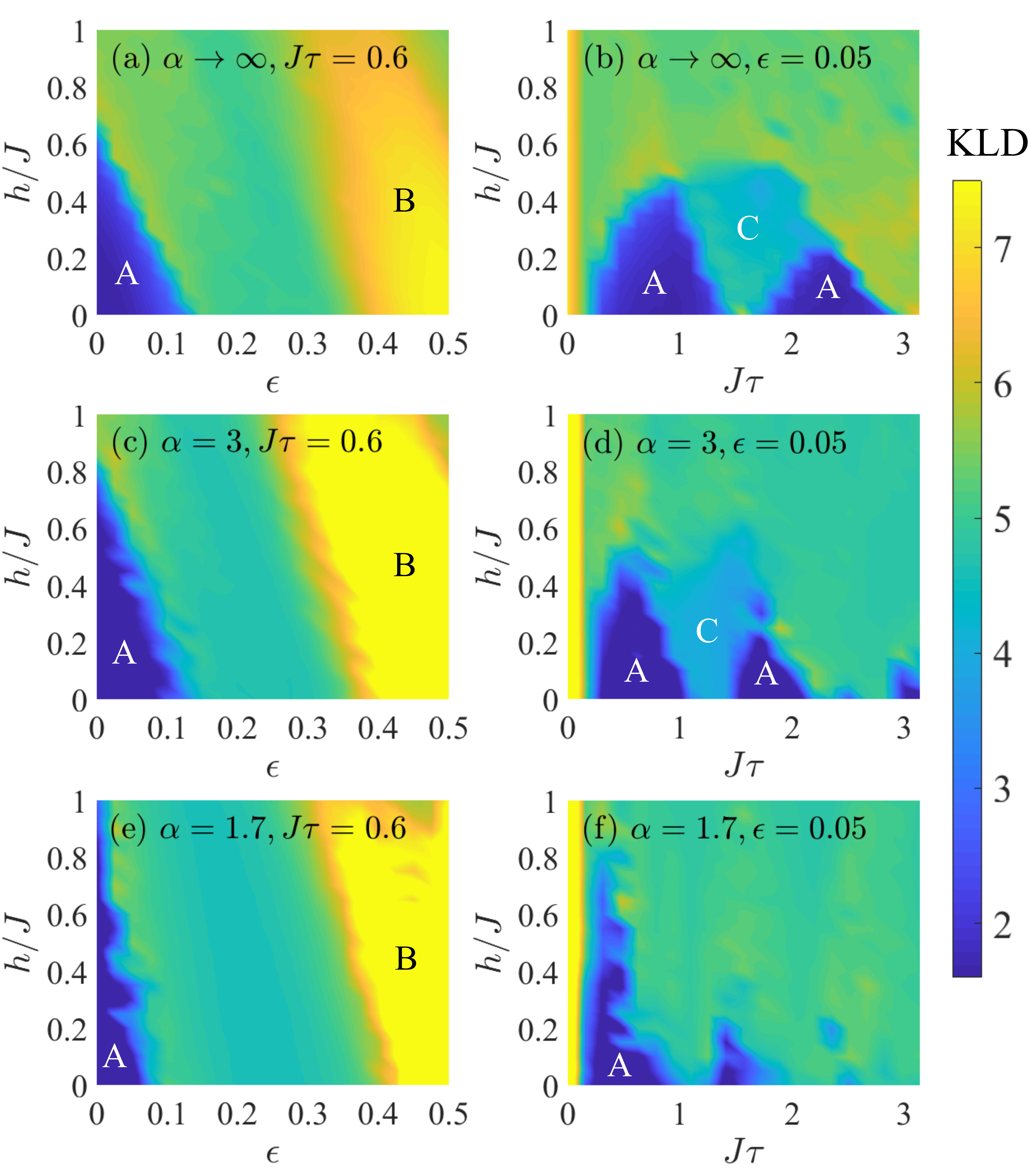}
	\caption{ Color map of the KLD in the $\epsilon-h/J$ and $J\tau-h/J$ parameter space for $N=14$. The top row corresponds to the nearest neighbour interacting case ($\alpha\rightarrow\infty$) and the interaction range increases down the row. A labels the DTC phase, B labels the regime in which the system oscillates with the drive, and C labels the regime where the drive hits the second harmonic of the system.}\label{fig6}
\end{figure}

%\begin{figure*}[t]\centering
%	\includegraphics[width=14cm]{figS1.pdf}
%	\caption{\label{figS1} Fourier spectrum of the stroboscopic magnetization in the $x$ direction with $N=14,h=0,J\tau=0.6$ for $\epsilon$ from three different phases. The Fourier transform is performed over 1000 periods in the main plots. (a) $\epsilon=0.1$ and the system is in DTC phase. Inset shows a zoom-in of the main peaks around $\omega\tau=\pi$. The Fourier transform is carried out over $10^5$ periods here. The two main peaks are separated by $(\epsilon/\epsilon^*)^{m_aN}$ in this phase. (b) $\epsilon=0.25$ and there is no prominent peak observed. (c) $\epsilon=0.5$ and the system oscillate with the drive. One prominent peak is observed at $\omega=0$. The spectrum is folded into $[-\pi,\pi)$ for better visualization of the main peak. }
%\end{figure*}

To validate the above picture, we calculate the time evolution using exact diagonalization. The corresponding driving frequency is $\omega_d=2\pi/\tau\approx 10.5J$, while the energy gap of $H_0$ is $\Delta E=2J$. In Fig. \ref{fig2}(c), we plot the splitting $\ln(\delta\omega)$ as a function of $\ln(\epsilon)$ for various system's sizes $N$. The data for each $N$ is fitted linearly, i.e., $\ln(\delta\omega) = b(N)+a(N)\ln(\epsilon)$. As shown in Fig. \ref{fig2}(d), $a(N)$ and $b(N)$ are approximately linearly dependent on $N$ with the slopes $m_a\approx0.88$ and $m_b\approx1.33$, respectively. This linear dependence agrees well with the perturbation theory which predicts $m_a=1$. In the limit $N\to\infty$, the splitting $\delta\omega\approx (\epsilon/\epsilon^*)^{m_aN}$ goes to zero when $\epsilon<\epsilon^*=\exp(-m_b/m_a)\approx 0.22$ and diverges otherwise.

To analyze the full spectrum including side peaks, in Fig. \ref{fig3}(a) we plot a color map of the Fourier spectrum as a function of $\epsilon$. The spectrum is divided into three regimes:  (1) $\epsilon\lesssim 0.14$ where there are two main peaks separated by $(\epsilon/\epsilon^*)^{m_aN}$ around $\omega\tau=\pi$, (2) $0.14\lesssim\epsilon\lesssim 0.35$ where there is no prominent peak, and (3) $0.35 \lesssim \epsilon< 0.5$ where there is one prominent peak at $\omega=0$. The corresponding Fourier spectra with values of $\epsilon$ taken from each of the regime are shown in Figs. \ref{fig3}(g)- \ref{fig3}(i).

To quantify the transitions, we calculate the Kullback-Leibler (KL) divergence  defined as \cite{Kullback1951}
%\texttt{\bea
%\textrm{KLD}=\sum_{\omega} A_{\omega}\ln\left(\frac{A_{\omega}}{A_{\omega}^{\textrm{ref}}}\right),
%\label{eq:kl}
%\eea}
\texttt{\bea
	\textrm{KLD}=\sum_{\omega} A_{\omega}\ln\left(A_{\omega}/A_{\omega}^{\textrm{ref}}\right),
	\label{eq:kl}
	\eea}
where $A_{\omega}$ is the Fourier spectrum of $m^x(n)$ and $A_{\omega}^{\textrm{ref}}$ is the Fourier spectrum of a perfect cosine function with $\omega_f\tau=\pi$ \cite{kl}. The sum of the Fourier spectra are normalized to one, i.e. $\sum_{\omega}A_{\omega}=\sum_{\omega}A_{\omega}^{\textrm{ref}}=1$. Physically, the KLD measures how the Fourier spectrum $A_{\omega}$ is different from $A_{\omega}^{\textrm{ref}}$ which signatures a perfect DTC.  Figure \ref{fig3}(d) shows the KLD as a function of $\epsilon$. The KLD shows distinct behaviour in the three regimes mentioned above, as expected. The dynamics of the three cases can be understood by considering three limiting cases. (1) When $\epsilon=0$, there is one prominent peak at $\omega\tau=\pi$ as discussed earlier. (2) When $\epsilon=0.25$, the kick operator rotates $\ket{\underline{R}}$ to $\prod_i (\ket{\rightarrow}_i+\ket{\leftarrow}_i)/\sqrt{2}$, hence maximizing the overlap with the excited states. Thus the Fourier spectrum shows no prominent peak. (3) When $\epsilon=0.5$, the kick operator is turned off and the state does not evolve. Hence, the Fourier spectrum has a prominent peak at $\omega=0$.

In Fig. \ref{fig3}(b), we plot the Fourier spectrum as a function of $J\tau$. The spectrum shows five different regimes as $J\tau$ is varied from zero to $\pi$. The corresponding transitions, labeled by $J\tau_q^*$ with $q\in\{1,2,3,4\}$, are captured by the KLD, as depicted in Fig. \ref{fig3}(e). These transitions can be understood as follows. In the limit of $J\tau=0$, as shown in Fig. \ref{fig2}(a), the spectrum displays two main peaks separated by $4\pi\epsilon$. When $J\tau$ is increased, these two peaks create a beating effect where the envelope oscillates over the period $\tau/\epsilon$. The kick operator creates excitations that oscillate on the timescale of $2\pi/\Delta E=\pi/J$. The first transition happens when these two timescales become comparable, i.e., $J\tau^*_1=\epsilon \pi \approx 0.4$. This approximated value agrees with the transition shown in Figs. \ref{fig3}(b) and \ref{fig3}(e). As $J\tau^*_1<J\tau<J\tau^*_2$, the drive is off-resonant with $\Delta E$ leading to the DTC as discussed earlier. At $J\tau=0.5\pi$, the drive hits the second harmonic ($\omega_d\sim 2\Delta E$) of the system. In this regime, after two driving periods, the excitations gain the phase of $2\Delta E\tau=2\pi$. Hence, in the rotating frame that oscillates with the period $2\tau$, the system will behave as if $J\tau=0$, leading to two peaks at $\pi\pm 2\epsilon\pi$. When moving back to the original frame that oscillates with the period $\tau$, there is an extra peak at $\pi$. The phase boundaries can be calculated as $J\tau^*_2=0.5\pi-\epsilon\pi$ and $J\tau^*_3=0.5\pi+\epsilon\pi$. At $J\tau=\pi$, the drive hits the first harmonic of the system. The excitations gain the phase of $2\pi$ after one driving period. Hence, the situation is the same as $J\tau=0$. The phase boundary is $J\tau^*_4=\pi-\epsilon\pi$. At $J\tau^*_3<J\tau<J\tau^*_4$, the drive is off-resonance with the first and the second harmonic of the system, leading to the DTC.

In Fig. \ref{fig3}(c), we plot the Fourier spectrum as a function of $h/J$. The spectrum shows a transition at $h^*/J\approx 0.5$ which also appears in the corresponding KLD plot in Fig. \ref{fig3}(f). At $h/J>h^*/J$, we observe that the splitting of the main peak grows linearly with $h/J$ with the rate $2\tan\theta\sim 0.40$. This splitting can be understood as follows. In the limit $h/J\gg 1$, the magnetic field dominates the spin-spin interactions and $[U_0,K_{\phi}]\simeq 0$. Hence, the system evolves with the approximated operator $K_{0.5\pi-\epsilon\pi-h\tau}=K_{\pi(0.5-\epsilon')}$, where $\epsilon'=\epsilon+h\tau/\pi$. Hence the splitting rate is $2\delta\epsilon'/\delta (h/J)=2\tau/\pi\sim 0.38$ ($J\tau=0.6$), in agreement with the splitting observed in Fig. \ref{fig3}(c). This relation also holds for other values of $J\tau$ as shown in Fig. \ref{fig4}.

We also consider an initial state prepared from one of the ferromagnetic ground states of $H_0$ in Eq. (\ref{eq:H_LRI}) with $0<h/J<1$. In experiment, such a state can be prepared by cooling the system in the presence of a strong magnetic field in the $x$ direction at two ends of the chain. As we can see from Fig. \ref{fig5}, the period doubling in the stroboscopic magnetization remains robust and persists in a large system size \cite{SI_DTC}.

For general values of the driving parameters $\epsilon,J\tau$ and $h/J$, the approximate phase boundaries of DTC are captured by the KLD as shown in Figs. \ref{fig6}(a) and \ref{fig6}(b). The DTC phases are stable up to $h/J\sim0.6$. For $h/J\ne 0$, the energy spectrum becomes dispersive and bands are formed. In particular, the energy span of the second band increases as $h/J$ increases. This results in the wedge-like shape of phase C in Fig. \ref{fig6}(b). Moreover, the first DTC phase (on the left) is more robust than the second DTC phase (on the right). This is because in order to get out of the first DTC phase, resonance to the states in the second band is required and is more difficult to achieve than populating the states in the first band (which melts the second DTC) since a higher order perturbation in the kick operator is involved.

%%%%%%%%%%%%%%%%%%%%%

\section{The effect of long-range interactions} 

\begin{figure*}[t]\centering
	\includegraphics[width=14cm]{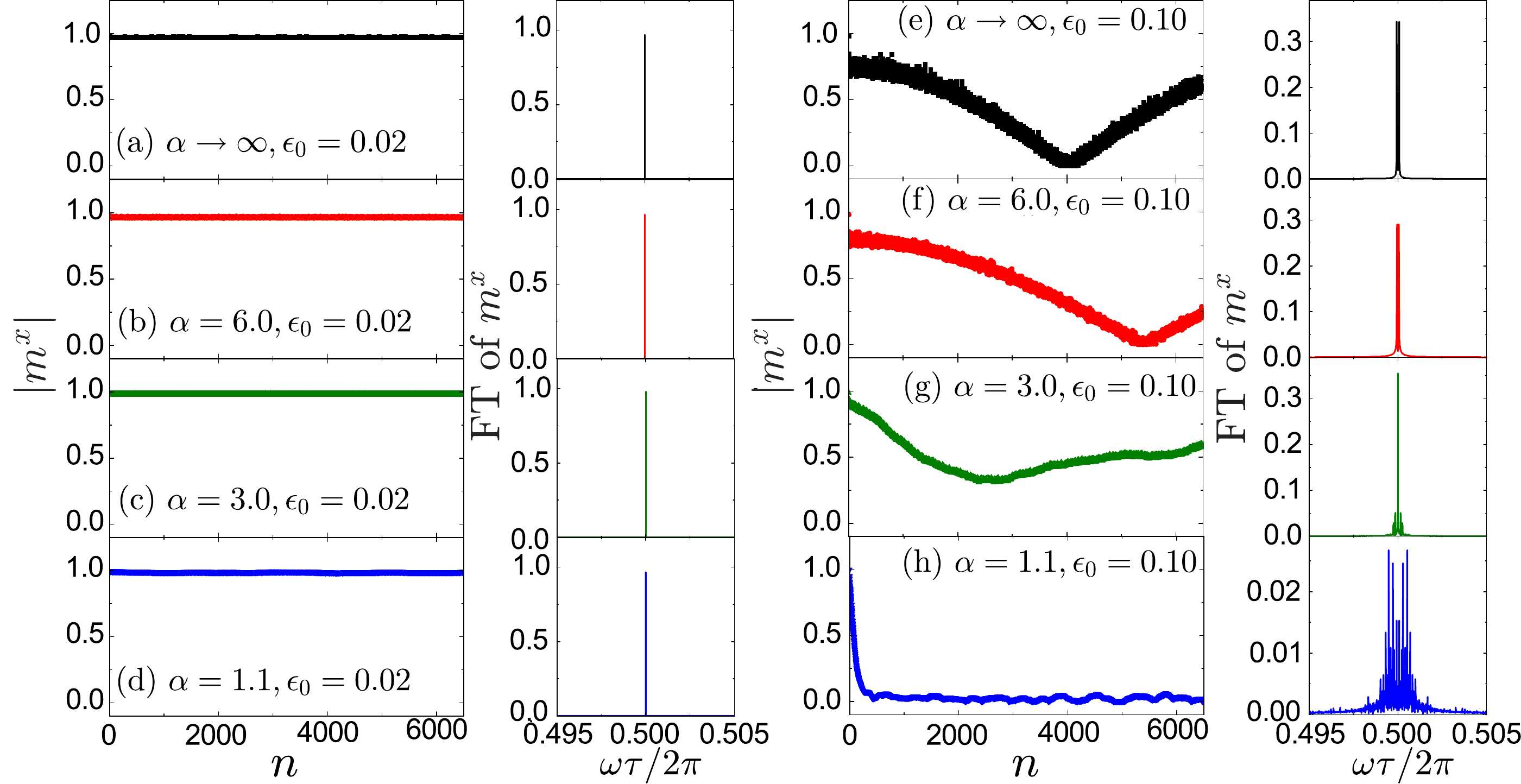}
	\caption{\label{fig7} Stroboscopic magnetization as a function of time and the corresponding Fourier spectrum for various $\alpha$ and $\epsilon_0$. Here $N=14,h/J=0.32,J\tau=0.6$.}
\end{figure*}

Now let us consider the effect of the range of interaction in stabilizing the DTC. The Hamiltonian is modified to $H_0=-J\sum_{i<j}\sigma^x_i\sigma^x_j/|i-j|^{\alpha}-h\sum_i\sigma^z_i$, with $\alpha$ characterizing the interaction range \cite{Jaschke2017,ZZhu2018}. Figures \ref{fig6}(c)-\ref{fig6}(f) show the phase diagrams for various values of $\alpha$ in the $(h/J)-\epsilon$ and $(h/J)-J\tau$ planes respectively. Upon decreasing $\alpha$ (increasing the range of interactions), the DTC phase becomes more robust to the perturbations in the external transverse field (left column of Fig. \ref{fig6}). The long-range interaction helps to maintain the system in the symmetry broken state with a finite magnetization in the $x$ direction as well as stabilizing the DTC against perturbations in the imperfection of the spin-flip $\epsilon$. On the other hand, the two DTC phases observed in the nearest neighbour interacting case in the $J\tau$ parameter space shrink upon introducing long range interactions. To understand this, let us consider the limiting case $h=0$. In the presence of long-range interactions, spin flips at different sites have different energies. This results in a broadening of the energy spectrum and increasing the probability of populating the excited states. Hence, the stability of the DTC phase decreases.

We further check the stability of the DTC by taking the initial state as one of the ground states of $H_0$ with $h/J=0.32$ and introducing noises in the kick by setting $\epsilon$ as a random variable, i.e. $K_{\phi}=\exp\left[{-i\sum_i\phi_i\sigma_i^z}\right]$, where $\phi_i=\pi(1/2-\epsilon_i)$ and $\epsilon_i$ is drawn from a uniform distribution $[0,\epsilon_0]$. The result is shown in Fig. \ref{fig7}. We can see from the left panel of the plot, the period doubling persists for a small $\epsilon_0$ and this further supports the presence of the DTC without fine-tuning the Hamiltonian parameters. As $\epsilon_0$ becomes larger, the DTC becomes less stable and the effect of noise is more pronounce in the case of smaller $\alpha$ as show in the right panel of Fig. \ref{fig7}.

%For $\alpha>3$, increasing the interaction range (decreasing $\alpha$) tends to stabilize the time crystal. However, further decreasing $\alpha$ beyond this threshold does not help to stabilize the DTC, and can even destabilize the time crystal. %The latter seemingly contradicts recent results obtained for the LMG model where the long range was suggested to help the DTC \cite{Russomanno2017}. The approach there assumed the extreme case of infinite range in which the Hamiltonian can be mapped to a single particle problem which effectively behaves as a classical top. For the intermediate range of interactions, which we study here, the semi-classical approach does not fully apply, and thus one should not expect the latter expectation to hold as our numerics confirm.

%\begin{figure}[t]
%  \centering
%  \includegraphics[width=9cm]{fig4.pdf}
%  \caption{\textbf{DTCs with long-range interactions. } Color maps of the KLD with various driving parameters for $N=8$ and the effect of interaction range on the DTC phase (indicated by the blue regions) boundaries.}\label{fig4}
%\end{figure}

\section{Conclusions}

We showed the possibility to observe a stable DTC in an Ising spin system in the absence of disorder subjected only to a periodic drive without the need for any other Hamiltonian control \cite{SI_DTC}. The simplicity of the global driving protocol should trigger further theoretical and experimental studies in this direction. Among future works, one could consider the effect of the shape of the periodic drive with Gaussian with finite lifetime instead of a delta. The possibility to observe such behavior for different spin Hamiltonians and the dimensionality dependence might be of interest as well.

\section*{Acknowledgements}
We thank J.I. Cirac for fruitful discussions and support from the National Research Foundation and the Ministry of Education of Singapore.
This research was also partially funded by Polisimulator project co-financed by Greece and the EU Regional Development Fund, the European Research Council under the European Union's Seventh Framework Programme (FP7/2007-2013)/ERCGrant Agreement No.319286 Q-MAC. D.J. acknowledges support from the EPSRC under Grant Nos. EP/K038311/1, EP/P01058X/1 and EP/P009565/1.

\pagebreak
\widetext
\begin{center}
	\textbf{\large Supplemental Material}
\end{center}
\setcounter{equation}{0}
\setcounter{figure}{0}
\setcounter{table}{0}
\setcounter{section}{0}
\makeatletter
\renewcommand{\theequation}{S\arabic{equation}}
\renewcommand{\thefigure}{S\arabic{figure}}
\renewcommand{\bibnumfmt}[1]{[S#1]}
\renewcommand{\citenumfont}[1]{S#1}

In this Supplemental Material, we derive the stroboscopic magnetization to the lowest order of $\epsilon$ from perturbation theory in section I.  In section II, the eigenspectrum of the time evolution operator and its relation to the existence of DTC in the nearest-neighbour Ising model is studied.

\section{Perturbative treatment in the infinite-range interacting case}
In this section, we expand the kick operator in terms of $\epsilon$ and show that the first order term gives rise to a main central peak and two side peaks of order $~\epsilon^2$ in the Fourier spectrum.

The kick operator up to the first order in $\epsilon\pi$ is given by
\bea
U_{\rm{kick}}=\Keps\Kpi=\sum_{\mu=0}^{\infty}(i\pi\epsilon S^z)^{\mu}\Kpi,
\eea
with short hand notation $S^z=\sum_i\sigma^z$, and $K_{\theta}=\exp\left[-i\theta\sum_i\sigma_i^z\right]$. Take the initial state $\ket{0}=\ket{\underline{R}}=|\rightarrow,\rightarrow,\rightarrow,...,\rightarrow\rangle$. At time $t=\tau$, the system is in the state
\bea
\ket{1}=U\ket{0}=[\Kpi+i\pi\epsilon\Kpi S^z]\ket{0}.
\eea
The state $S^z\ket{\underline{L}}$ where $|\underline{L}\rangle\equiv|\leftarrow,\leftarrow,\leftarrow,...,\leftarrow\rangle$ is in general not an eigenstate of $H_0$, i.e. $S^z\ket{\underline{L}}=\sum_{\mu}c_{\mu}\ket{\mu}$ with coefficients $c_{\mu}$.
At $t=2\tau$, the system is in the state
\bea
\ket{2}=U\ket{1}=[\Kpi^2+i\pi\epsilon(S^z\Kpi^2+\Kpi U_0 S^z\Kpi]\ket{0},
\eea
where $U_0=\exp[-iH_0\tau]$.
At $t=3\tau$, the system is in the state
\bea
\ket{3}=U\ket{2}=[\Kpi^3+i\pi\epsilon(\Kpi^3+\Kpi U_0 \Kpi^2+(\Kpi U_0)^2S^z]\ket{0},
\eea
and at $t=4\tau$,
\bea
\ket{4}=U\ket{3}=[\Kpi^4+i\pi\epsilon(\Kpi^4+\Kpi U_0 \Kpi^3+(\Kpi U_0)^2\Kpi^2+(\Kpi U_0)^3\Kpi)S^z]\ket{0}.
\eea
In general, at $t=n\tau$, one obtains
\bea
\ket{n}=\left[\Kpi^n+i\pi\epsilon\left(\sum_{\nu=0}^{n-1}(\Kpi U_0)^{\nu}\Kpi^{n-\nu}\right)S^z\right]\ket{0}+\mathcal{O}(\epsilon^2).
\label{eq:psin}
\eea

For the LMG model where the interaction is of infinite range, the Hamiltonian reads
\bea
H_{\rm{LMG}}=-\frac{1}{4N}\sum_{i,j}\sigma^x_i\sigma^x_j-\frac{h}{2}\sum_i\sigma^z_i.
\eea
The prefactor $1/N$ is to ensure the energy is extensive in the thermodynamic limit. Using the total spin operator $S^{\kappa}=\sum_i\sigma^{\kappa}_i/2$ (with $\kappa=x,y,z$) obeying $[S^z, J^{\pm}]=\pm J^{\pm}$ and $[J^+,J^-]=2J^z$, the Hamiltonian can be rewritten as a single particle model
\bea
H_{\rm{LMG}}=-\frac{1}{N}(S^x)^2-hS^z.
\eea
This Hamiltonian commutes with the total spin $\textbf{S}^2$, thus conserving angular momentum, and with $\Kpi$, corresponds to the parity symmetry. Thus, $H_{\rm{LMG}}$ can be diagonalized in each sector $S\in[0,N/2]$ separately. In the subspace of $S=N/2$ which contains the ground state, the eigen-basis is given by the Dicke states $\ket{S,M}$ where $M\in[-S,-S+1,\cdots,S]$. For the ferromagnetic interaction that we are considering here, the ground state is $\ket{S,S}$ and $\ket{S,-S}$.

Since $[H_0,\Kpi]=0$ for the LMG model, Eq. \ref{eq:psin} simplifies to
\bea
\ket{n}=\Kpi^n\left[1+i\pi\epsilon\sum_{\nu=0}^{n-1}U_0^{\nu}S^z\right]\ket{0}+\mathcal{O}(\epsilon^2),
\eea
with $\ket{0}=\ket{S,S}$. The action of $S^z$ generates a lower eigenstates of the Dicke ladder, i.e. $S^z\ket{S,S}=\sqrt{N}\ket{S,S-1}$. Using $H_{\rm{LMG}}\ket{S,S}=0$ and $H_{\rm{LMG}}\ket{S,S-1}=\omega_1\ket{S,S-1}$, we find
\bea
\ket{n}&=&\Kpi^n\ket{S,S}+i\pi\epsilon\sqrt{N}\sum_{\nu=0}^{n-1}(e^{-i\omega_1\tau})^{\nu}\Kpi^2\ket{S,S-1},\\ \nonumber
&=&\Kpi^n\ket{S,S}+i\pi\epsilon\sqrt{N}\chi_n\Kpi^2\ket{S,S-1},
\eea
where
\bea
\chi_n=\frac{1-e^{-i\omega_1n\tau}}{1-e^{-i\omega_1\tau}}.
\eea
To calculate the magnetization
\bea
m^z_n=\frac{1}{N}\mele{n}{S^x}{n},
\eea
we start with
\bea
S^x\ket{n}=(-1)^nS\ket{S,S}+i\epsilon\pi\chi_n(-1)^n\sqrt{N}(S-1)\ket{S,S-1}
\eea
using the fact that the Dicke states are eigenstates of $S^x$. For large $N$, we can approximate $S-1\approx S=N/2$. We get
\bea
m^x_n=(-1)^n\left[1-\epsilon\pi^2|\chi_n|^2N\right].
\eea
Using $(-1)^n=\cos(\pi n)$ and
\bea
|\chi_n|^2=\frac{1-\cos(\omega_1\tau n)}{1-\cos(\omega_1\tau)},
\eea
we find
\bea
m^x_n=\left[1-\frac{\epsilon^2\pi^2N}{1-\cos(\omega_1\tau)}\right]+\frac{\epsilon^2\pi^2N}{1-\cos(\omega_1\tau)}\frac{1}{2}[\cos(n\pi-n\omega_1\tau)+\cos(n\tau+n\omega_l\tau)].
\eea
The Fourier transform of the above equation gives three peaks at 0 and $\omega_l\pi$ with heights depending on the prefactor of the cosine functions which in turn depends on $\tau$ and $\epsilon$.

\section{Spectrum of the time evolution operator in the time crystal}

To have a stable time crystal phase, Ref. \cite{Keyserlingk2016_S} showed that the spectrum of the time evolution operator (Floquet operator) has to have a particular structure. Unless otherwise specified, we consider $J=1$ in this section. Any eigenstate of the Floquet operator with quasi-energy $\mu_{\alpha}$ needs to have a partner with quasi-energy $\mu_{\alpha}+\pi/\tau$ (where $\tau$ is the driving period), a.k.a $\pi$ spectral pairing.

To understand this, let's recall the Floquet unitary operator of our system
\bea
U=K_{-\epsilon\pi}K_{\pi/2}U_0,
\eea
where $U_0=\exp\left[{-iH_0\tau}\right]$ is the free evolution operator and $K_{\phi}=\exp\left[-i\phi\sum_i\sigma_i^z\right]$. The eigensystem of the Floquet operator is usually written as
\bea
U(\tau)\ket{\alpha}=e^{-i\mu_{\alpha}\tau}\ket{\alpha},
\eea
and $\mu_{\alpha}$ is known as the quasi-energy.

Note that the parity operator $P=\prod_i\sigma^z_i$ commutes with $H_0$ as well as $U$. Both the eigen-energy $E_s$ of $H_0$ and the parity eigenvalues $p=\pm 1$ are good quantum numbers. For illustration, consider the case of $h=0$ and perfect spin flip $\epsilon=0$, the eigenstates of the Floquet operator can be expressed in the form of
\bea
\ket{\pm}\sim\ket{E_s,p=\pm 1}=\frac{1}{\sqrt{2}}\left(\ket{\{\sigma^x_s\}}+\ket{\{\overline{\sigma^x_s}\}} \right),
\eea
where $\ket{\{\sigma^x_s\}}$ and $\ket{\{\overline{\sigma^x_s}\}}$ are the $\mathbb{Z}_2$ symmetry-breaking states of $H_0$. We have
\[ U\ket{\pm}=
\begin{cases}
& \pm e^{-iE_s\tau}\ket{\pm} \quad\text{for}\quad N=4m \\
& \mp e^{-iE_s\tau}\ket{\pm} \quad\text{for}\quad N=4m+2 \\
\end{cases}
\]
where $N$ is the system size and $m$ is a positive integer. In either case, we have a pair of eigenstates $\{\ket{+},\ket{-}\}$ having an eigenvalue of opposite sign. This translates into a $\pi/\tau$ difference in the quasi-energy. If one prepares the initial state as a superposition of the $\ket{+}$ and $\ket{-}$, that is a symmetry broken state of $H_0$, it will undergo Rabi oscillation with a frequency $\pi/\tau$. More explicitly,
\bea
\ket{\Psi(0)}&=&\frac{1}{\sqrt{2}}(\ket{+}+\ket{-})\\
\ket{\Psi(n\tau)}&=&U_{f0}^n\ket{\Psi(0)}=\frac{1}{\sqrt{2}}(e^{-i\mu_+n\tau}\ket{+}+e^{-i\mu_-n\tau}\ket{-}).
\eea
If $\mu_-=\mu_++\pi/\tau$,
\[ \ket{\Psi(n\tau)}=
\begin{cases}
& \frac{1}{\sqrt{2}}e^{-i\mu_+n\tau}(\ket{+}-\ket{-}) \quad\text{for odd $n$} \\
& \frac{1}{\sqrt{2}}e^{-i\mu_+n\tau}(\ket{+}+\ket{-})=e^{-i\mu_+n\tau}\ket{\Psi(0)} \quad\text{for even $n$} \\
\end{cases}
\]
and so the observable returns to itself every $2\tau$.

\begin{figure}[H]\centering
	\includegraphics[width=18cm]{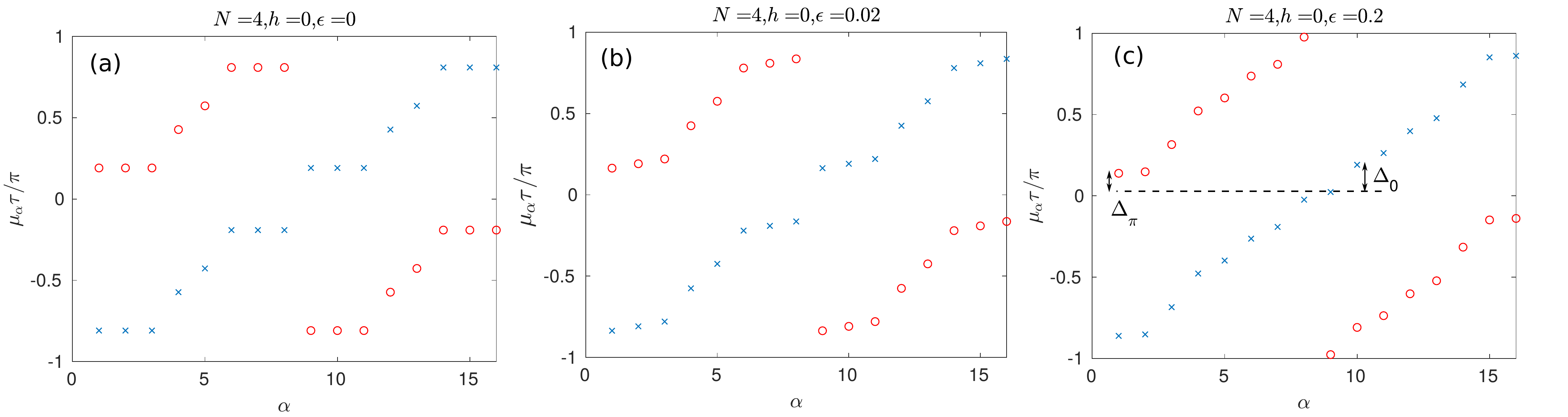}
	\caption{\label{figS4} Blue crosses show the Floquet spectrum for various $\epsilon$ for $N=4$, $h=0$ and $\tau=0.6$. The red circles are the $\pm\pi/\tau$ shift of the quasi-energies. For $\epsilon=0$, each of the blue cross with quasi-energy $\mu_{\alpha}$ has a red circle that corresponds to an eigenstate with quasienergy $\mu_{\alpha'}\pm\pi/\tau$ to pair up. For a larger value of $\epsilon=0.2$, the Floquet spectrum is significantly modified and the $\pi$ pairing of Floquet eigenstates is inhibited. }
\end{figure}

Figure \ref{figS4} illustrates the $\pi$ pairing in the Floquet spectrum for $N=4$ and $h=0$ for the NN Ising model. The Floquet spectrum is modified when $\epsilon\ne0$. Depending on the value of $\epsilon$ and how the energy gap between adjacent states and that of the even and odd parity states are modified, the pairing may be inhibited (like in Fig. \ref{figS4}(c)) and the period doubling in the observable does not persist.

More generally, Ref. \cite{Keyserlingk2016_S} and \cite{Russomanno2017_S} introduced a scheme to check for the $\pi$-spectral pairing by considering the quasi-energy gaps
\bea
\Delta^{(\alpha)}_0 =\mu_{\alpha+1}-\mu_{\alpha} \quad\text{and}\quad  \Delta^{(\alpha)}_{\pi} =\mu_{\alpha+\mathfrak{N}/2}-(\mu_{\alpha}+\pi/\tau),
\eea
for all $\alpha$'s, and $\mathfrak{N}$ is the dimension of the Hilbert space. If the system is a time crystal and there's $\pi$-spectral pairing, we expect $\avg{\log{\Delta_{\pi}}}$ to be much smaller than $\avg{\log{\Delta_{0}}}$. Moreover, in order to have the pairing in the thermodynamic limit, we need $\avg{\log{\Delta_{\pi}}}$ scales down faster than $\avg{\log{\Delta_{0}}}$ with the system size.

\begin{figure}[H]\centering
	\includegraphics[width=12cm]{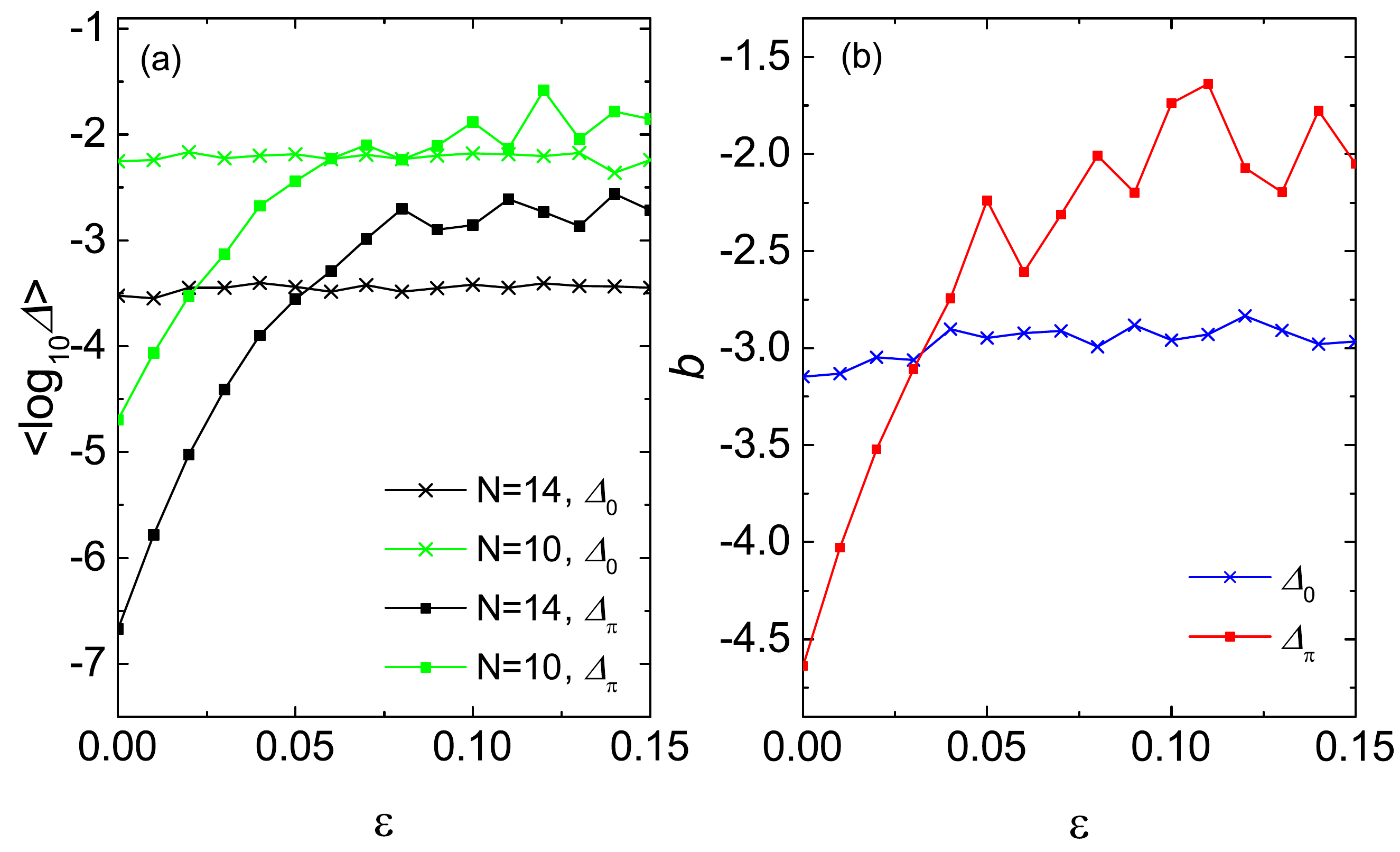}
	\caption{\label{figS5} (a) The adjacent quasienergy gap $\Delta_0$ and the even-odd parity states gap $\Delta_{\pi}$ as a function of $\epsilon$. (b) Scaling exponent of $\avg{\log{\Delta_{0/\pi}}}$ to the system size $N$ as a function of $\epsilon$. For $\epsilon\lesssim 0.05$, the $\pi$-spectral pairing is favorable. Here the NN Ising model for $h=0.32$ and $\tau=0.6$ is considered.}
\end{figure}

Figure \ref{figS5}(a) shows a plot of $\avg{\log{\Delta_{0/\pi}}}$ as a function of $\epsilon$ in the NN Ising model for $h=0.32,\tau=0.6$. For each of the $\epsilon$, we perform numerical linear fitting to
\bea
\avg{\log{\Delta_{0/\pi}}}=a+b\log N,
\eea
and plotted the slope as a function of $\epsilon$ in Fig. \ref{figS5}(b).

%\section{VI. With a noisy kicking protocol}
%In the main text, we considered a delta kick parameterized by a fixed variable $\epsilon$. One can also introduce noise in the kick by taking $\epsilon$ as a random variable, i.e. $U_{\rm{kick}}=\exp\left[{-i\sum_i\phi_i\sigma_i^z}\right]$, where $\phi_i=\pi(1/2-\epsilon_i)$ and $\epsilon_i$ is drawn from a uniform distribution $[0,\epsilon_0]$. The result is shown in Fig. \ref{figS6}. The DTC phase remain robust.
%
%\begin{figure}[H]\centering
%              \includegraphics[width=16cm]{figS6.pdf}
%              \caption{\label{figS6} Stroboscopic magnetization as a function of time and the corresponding Fourier spectrum for various $\alpha$ and $\epsilon_0$. Here $N=14,h/J=0.32,J\tau=0.6$.}
%\end{figure}


\begin{thebibliography}{}
\bibitem{kl}
	A small noise of order $10^{-5}$ is added to $A^{\rm ref}_{\omega}$ to avoid divergence caused by a zero in the denominator in Eq. (\ref{eq:kl}).
\end{thebibliography}
\end{document}